%% file: 0_paper.tex
\documentclass[a4paper,fleqn]{cas-dc}

\usepackage[numbers]{natbib}

\ExplSyntaxOn
\cs_gset:Npn \__first_footerline:
  { \group_begin: \small \sffamily \__short_authors: \group_end: }
\ExplSyntaxOff 

\def\tsc#1{\csdef{#1}{\textsc{\lowercase{#1}}\xspace}}
\tsc{WGM}
\tsc{QE}
\tsc{EP}
\tsc{PMS}
\tsc{BEC}
\tsc{DE}

\usepackage[dvipsnames]{xcolor}
\usepackage{graphicx}
\usepackage{textcomp}
\usepackage{xcolor}
\usepackage{subcaption}
\usepackage{colortbl}
\usepackage{enumerate}
\usepackage{algpseudocode}
\usepackage{enumitem}
\usepackage{graphics}
\usepackage{amsmath}

\usepackage{graphicx}
\usepackage{float}
\usepackage{placeins}

\usepackage{awesomebox} 
\usepackage{fontawesome5}
\usepackage{hhline}
\definecolor{findingBarColour}{gray}{0.4}
\definecolor{findingIconColour}{gray}{0.3}

\newcommand\revision[1]{#1}

\newcommand\example[1]{\textit{e.g., "#1"}}

\usepackage{tcolorbox}

\tcbuselibrary{skins}
\newtcolorbox{rqbox}[1]
{
  before skip=1em, 
  after skip=1em, 
  colframe = black!60,
  colback  = black!10,
  coltitle = black!90,  
  title    = \textbf{#1},
  hbox boxed title,
  enhanced,
  attach boxed title to top center={yshift=-3mm,yshifttext=-1mm},
  boxed title style={size=small, colback=black!30}
}

\usepackage{siunitx}
\newcommand{\commentout}[1]{}

\usepackage{booktabs}

\usepackage{amssymb}
\usepackage{graphicx}
\usepackage{textcomp}
\usepackage{xcolor}
\usepackage{subcaption}
\usepackage{colortbl}
\usepackage{enumerate}
\usepackage{algpseudocode}

\usepackage{color}
\usepackage{hyperref}
\usepackage{xcolor}

\usepackage{booktabs}
\usepackage{array}

\usepackage[utf8]{inputenc}

\usepackage{adjustbox} 

\usepackage{booktabs}
\usepackage{amsmath}
\usepackage{array}

\usepackage{tcolorbox}

\begin{document}
\let\WriteBookmarks\relax
\def\floatpagepagefraction{1}
\def\textpagefraction{.001}

\shorttitle{Robotics Meets Software Engineering: A First Look at the Robotics Discussions on StackOverflow}

\shortauthors{Kidwai et al.}

\title [mode = title]{Robotics Meets Software Engineering: A First Look at the Robotics Discussions on StackOverflow}                      



%

\author[1]{Hisham Kidwai}[]

\ead{kidwaih1@myumanitoba.ca}

\affiliation[1]{organization={SQM Research Lab, Department of Computer Science, University of Manitoba},
    city={Winnipeg},
    country={Canada}}

\affiliation[2]{organization={HCI Lab, Department of Computer Science, University of Manitoba},
    city={Winnipeg},
    country={Canada}}

\author[2]{Danika Passler Bates}


\ead{passlerd@myumanitoba.ca}





\author[1]{Sujana Islam Suhi}[
   ]
\ead{suhisi@myumanitoba.ca}

\author[1]{Md Nahidul Islam Opu}[
]
\ead{opumni@myumanitoba.ca}

\author[2]{James E. Young}[
   ]
\ead{young@cs.umanitoba.ca}

\author[1]{Shaiful Chowdhury}[
   ]
\ead{shaiful.chowdhury@umanitoba.ca}









\begin{abstract}
Robots can greatly enhance human capabilities, yet their development presents a range of challenges. This collaborative study, conducted by a team of software engineering and robotics researchers, seeks to identify the challenges encountered by robot developers by analyzing questions posted on StackOverflow. We created a filtered dataset of 500 robotics-related questions and examined their characteristics, comparing them with randomly selected questions from the platform. Our findings indicate that the small size of the robotics community limits the visibility of these questions, resulting in fewer responses. While the number of robotics questions has been steadily increasing, they remain less popular than the average question and answer on StackOverflow. This underscores the importance of research that focuses on the challenges faced by robotics practitioners.

Consequently, we conducted a thematic analysis of the 500 robotics questions to uncover common inquiry patterns. We identified 11 major themes, with questions about robot movement being the most frequent. Our analysis of yearly trends revealed that certain themes, such as \textit{Specifications}, were prominent from 2009 to 2014 but have since diminished in relevance. In contrast, themes like \textit{Moving}, \textit{Actuator}, and \textit{Remote} have consistently dominated discussions over the years. These findings suggest that challenges in robotics may vary over time.

Notably, the majority of robotics questions are framed as \textit{How} questions, rather than \textit{Why} or \textit{What} questions, revealing the lack of enough resources for the practitioners. These insights can help guide researchers and educators in developing effective and timely educational materials for robotics practitioners.
\end{abstract}



\begin{keywords}
StackOverflow \sep Robotics \sep Software Engineering
\end{keywords}

\maketitle

\input{1_intro}

\input{2_background}

\input{3_rw}

\input{4_methodology}

\input{5_results}

\input{6_discussion}

\input{7_conclusion}




\bibliographystyle{elsarticle-num}

\bibliography{paper, bg}





\end{document}

%% file: 1_intro.tex

\section{Introduction}
\label{sec:intro}
Robots are undergoing a revolution, increasingly integrating into various facets of our daily lives. Research highlights several benefits of interacting with robots, including enhanced motivation for exercise \cite{Fasola-1:2013}, \cite{Hebesberger-2:2016}, reduced loneliness, increased engagement in educational activities \cite{Hong-3:2016}, and emotional support \cite{Hung-4:2019}. Despite these advantages, the development of robots remains a complex challenge \cite{garcia2023software, zhang2018development}. Robots must interact with and react to objects and people in the real world. To cope with this, many robots still require additional support structures, such as human operators who remotely control the robots to perform tasks like interacting with others \cite{Logan-5:2019}. 

Software is integral to robot control, making software practitioners essential to the development process. Their responsibilities encompass implementing algorithms, ensuring real-time processing, and addressing bugs and performance issues. However, a recent workshop summary by leading experts in robotics and software engineering highlights that software development processes, techniques, and tools for robotics have remained largely unchanged for decades, impeding progress in the field~\cite{goues2024software}. While developing new techniques and tools is essential, the specific challenges faced by software practitioners—such as developers and testers—in creating robotic software remain largely underexplored~\cite{garcia2020robotics}.

In this study, we use developer Q\&A as an empirical lens to examine where robotics practitioners seek help while building robotic systems. Specifically, we analyze robotics-related questions on \textsc{StackOverflow}\footnote{https://stackoverflow.com (last accessed: 15-Aug-2024)}, a widely used programming Q\&A platform that has supported prior research on practitioner discussions in areas such as big data \cite{Bagherzadeh-big-data:2019}, security \cite{Xin-Li-security:2016}, off-topic discussions \cite{S.A.Chowdhury:2015}, machine learning \cite{Bangash-ml:2019}, and blockchain \cite{Wan-blockchain:2019}. Our objective is to characterize recurring questions and problem areas in robotics development and to relate these observed difficulties to aspects of system construction and design, such as component integration, configuration, and operational troubleshooting. The resulting insights are intended to inform future research directions and help educators and tool builders prioritize practical guidance and learning materials.


Overall, we examine the characteristics, themes, evolution, and underlying nature of the questions developers ask about robotics. This paper's contribution is based on an analysis of 500 carefully selected robotics-related questions, structured around the following four research questions.

\textbf{RQ1: What are the distinct characteristics of robotics-related questions?}
We demonstrate that the volume of robotics-related questions is increasing over time. Additionally, the lower number of views but comparatively higher number of answers to these questions suggests a smaller, yet highly engaged robotics community on  \textsc{StackOverflow}.

\textbf{RQ2: What are the most common themes in robotics discussions?}
 Through thematic analysis, we identified 11 major themes and 33 sub-themes in robotics-related questions. We observed that questions about robot \emph{Movement} are the most common (16.8\%), while those concerning generic \emph{Programming} are the least frequent (2.4\%). Notably, questions related to \emph{Task Management} issues are the most successful, receiving a higher number of accepted answers compared to other themes.

\textbf{RQ3: How do the themes of robotics questions evolve over time?}
Our investigation into the temporal evolution of robotics questions revealed that some themes, such as \emph{Specifications} (e.g., OS, or language selection) and \emph{Task Management} (e.g., predicting next command), are more prevalent in older questions, while themes like \emph{Programming} and \emph{Error} have become more common in recent years. The prevalence of certain themes, such as \textit{Moving} and \textit{Remote}, has remained consistent throughout our data collection period from 2009 to 2024.

\textbf{RQ4:  What are the distributions of different types and cognitive load associated with robotics questions?}
To assess the types and cognitive load~\cite{paas1994instructional,van2005cognitive} associated with robotics questions, we categorized the selected 500 questions into four types: \emph{what}, \emph{why}, \emph{how}, and \emph{other}. We found that half of these questions are \emph{how} questions, which implies the lack of resources available to robotics developers~\cite{Uddin:2021}. The prevalence of \emph{how} questions also indicates that answering most robotics questions demands a moderate level of cognitive effort. 




%% file: 2_background.tex
\section{Background}\label{sec:background}

Robots are embodied, programmable systems that sense and act in the physical world via integrated hardware (e.g., sensors, actuators, power, and computation) and software (e.g., estimation, control, planning), often operating with partial autonomy in dynamic environments~\cite{haidegger2021taxonomy}. Across application domains, robots are commonly instantiated as fixed-base manipulators (e.g., industrial arms), mobile ground platforms, and specialized mobile systems such as unmanned aerial and underwater vehicles~\cite{kendoul2012survey,wynn2014autonomous}.

Robotic systems are typically engineered as layered stacks that bridge low-level device access (drivers and real-time control) to higher-level autonomy (perception, reasoning, and task execution) \cite{kortenkamp2016robotic,siegwart2011introduction,ahmad2016software}. At the lowest level, feedback control stabilizes robot dynamics and enforces safety and performance constraints; at higher levels, state estimation and perception transform noisy sensor streams into representations suitable for decision-making \cite{thrun2002probabilistic,siegwart2011introduction}. To manage integration complexity and promote reuse, robotics middleware and frameworks provide standardized communication, naming, and packaging mechanisms that enable modular composition of components (e.g., ROS, Player, OROCOS) \cite{quigley2009ros,gerkey2001most,bruyninckx2001open}.

Robotic architectures are often described by how responsibility is allocated between deliberative and reactive control across three fundamental paradigms. The hierarchical paradigm is purely deliberative and emphasizes explicit planning and reasoning to select actions over longer horizons, while the reactive paradigm prioritizes low-latency sensor-to-actuator mappings to cope with time-critical interaction \cite{ingrand2017deliberation,brooks2003robust}. Hybrid architectures combine these strengths by separating fast reactive control from slower planning and sequencing, as exemplified by three-layer and three-tier (3T) architectures \cite{gat1998three,peter1997experiences}. More recently, service-oriented and cloud-enabled perspectives extend modularity beyond a single platform by treating computation, data, and even learned capabilities as network-accessible resources, introducing new opportunities and new dependability challenges for robot autonomy at scale \cite{kehoe2015survey,ahmad2016software,goues2024software}.

%% file: 3_rw.tex
\section{Related Works}
\label{sec:relatedWork}

\textsc{StackOverflow} is the flagship site of the Stack Exchange Network, hosting millions of users and a large corpus of questions and answers, highlighting its pivotal role for software practitioners. This vast repository of information has significantly impacted various domains of software engineering research~\cite{tanzil2024systematic} including software architecture and design~\cite{bi2021mining}, social aspects of software engineering~\cite{gantayat2015synergy}, API documentation~\cite{treude2016augmenting}, machine learning-based classifiers~\cite{S.A.Chowdhury:2015}, and recommender systems~\cite{diamantopoulos2015employing}, among others. 
Similar to our study, several investigations utilizing \textsc{StackOverflow} data have concentrated on analyzing questions and answers to gain insights into practitioners' discussions on specific topics. These studies aim to guide future research and assist educators in enhancing their training materials. 

Pinto \textit{et al.}~\cite{Pinto:2014} investigated developers' concerns about software energy consumption and their suggestions for improvement. Their study revealed that energy discussion among developers can be classified into five major themes. They also found seven significant causes of energy leaks. Our work is methodologically inspired by this study, particularly in its use of thematic analysis and popularity metrics; however, we extend this line of research by examining the temporal evolution of topics and incorporating cognitive load analysis to better understand the nature and complexity of practitioners’ inquiries in the robotics domain.

Yang \textit{et al.}~\cite{Xin-Li-security:2016} analyzed security-related posts using Latent Dirichlet Allocation (LDA) to uncover the types of security questions developers ask. In those questions, they found five major themes, including web and mobile security. 
Bangash \textit{et al.}~\cite{Bangash-ml:2019} identified 50 important topics among all the posted questions related to machine learning. They also found that those 50 topics can be merged into five major categories, including framework and algorithm. A similar study was conducted by Han \textit{et al.}~\cite{Han:2020} who focused on questions related to deep learning frameworks. Our work shares the goal of characterizing practitioner concerns but differs by employing manual thematic analysis and extending the analysis to examine the temporal evolution of topics and the cognitive load implied by different types of robotics questions.

Uddin \textit{et al.}~\cite{Uddin:2021} analyzed IoT-related questions and discovered 40 important topics that can be further merged into four high-level categories. Rosen \textit{et al.}~\cite{Rosen:2016} investigated questions specifically asked by mobile developers. Similarly, Abdellatif \textit{et al.}~\cite{Abdellatif:2020} utilized \textsc{StackOverflow} to identify challenges in chatbot development. Their findings revealed that posts related to chatbot creation and integration into websites attract the greatest interest from chatbot developers. Kochhar \textit{et al.}~\cite{kochhar:2016} studied questions related to software testing and found that most of the questions fall within the test framework or database category. Similar other studies have focused on refactoring~\cite{peruma2022refactor}, web services~\cite{mahmood2023empirical}, non-functional requirements~\cite{zou2015non}, and concurrency~\cite{ahmed2018concurrency}. Our work aligns with these studies in its research questions and analytical dimensions but differs in its focus on robotics, a domain characterized by tighter software–hardware coupling.


Beyond analyses grounded in \textsc{StackOverflow} data, robotics software engineering research has extensively examined the intrinsic challenges of developing robotic systems. Quigley \textit{et al.}~\cite{quigley2009ros} and Cousins \textit{et al.}~\cite{cousins2010sharing} report that systems such as ROS support reuse and hardware abstraction, while simultaneously introducing additional complexity through distributed execution and configuration management. Albonico \textit{et al.}~\cite{albonico2023software} present a systematic mapping study of software engineering research centered on ROS, characterizing ROS as a widely adopted robotics framework and synthesizing how topics such as quality assurance, testing, and maintainability have been investigated in ROS-focused work. Ahmed \textit{et al.}~\cite{ahmad2016software} conduct a systematic mapping of software architecture for robotic systems and identify eight thematic areas supporting operation, evolution, and development activities. From both industrial and academic perspectives, Garcia \textit{et al.}~\cite{garcia2020robotics} further identify limited adoption of software engineering best practices in robotics and persistent challenges related to software--hardware interoperability, modularity, and reuse. Dos Santos \textit{et al.}~\cite{dos2020preliminary} map software engineering research on robotic systems through a software-quality lens, cataloguing studies and associated quality attributes; they report that security is comparatively under-examined within this body of work. In contrast to these primarily survey- and interview-based studies, our work complements existing findings by grounding them in large-scale, real-world developer discourse. By analyzing robotics-related questions on \textsc{StackOverflow}, we capture organically occurring development challenges, misconceptions, and learning barriers as they arise during day-to-day practice.

Although numerous studies have utilized \textsc{StackOverflow} data across various domains, none have specifically examined robotics discussions, despite their significance within the community~\cite{goues2024software}. This paper seeks to bridge that gap by adopting methodologies inspired by related research and analyzing robotics-related questions. By complementing existing interview-based findings~\cite{garcia2020robotics}, we aim to provide deeper insights into the obstacles faced by robotics practitioners and contribute to a more comprehensive understanding of their needs.

%% file: 4_methodology.tex
\section{Data Collection}
\label{sec:methodology}
Our data collection combined both quantitative data (such as answer count, views, up-votes, etc.) and qualitative data (including titles and question bodies).
We first identified the tags necessary for collecting questions related to robotics from \textsc{StackOverflow}. Two authors, one specializing in software engineering and the other in robotics, collaborated to determine the most relevant tags. We ultimately selected two tags, \emph{robot} and \emph{robotics}. A question was considered related to robotics if it had at least one of these tags, and was therefore included in our dataset. The final dataset contains 1,402 questions: 87 tagged with \emph{robot}, 1,320 tagged with \emph{robotics}, and 5 tagged with both. We also considered additional tags like \emph{robotframework}, \emph{robotium}, and \emph{roboto}, but these resulted in too many false positives, as determined by our manual analysis. Even the two conservative tags we selected are not entirely free from false positives, albeit to a lesser extent, which we will discuss later. There is also a dedicated robotics channel within the \textsc{StackExchange} site\footnote{https://robotics.stackexchange.com (last accessed: 16-Aug-2024)} that can be used to extend our study if a classifier can be developed to filter out non-robotics questions. We discuss this in Section~\ref{sec:discussion} as a potential direction of future work.

We utilized the Stack Exchange Data Explorer (SEDE)\footnote{https://data.stackexchange.com/stackoverflow/query/new} to automate our data collection process. The dataset, extracted on April 3, 2024, encompasses data from August 1, 2009, to March 27, 2024. It includes all the essential information needed to address our four research questions, such as question titles, bodies, tags, views, and answers.





As noted earlier, our manual inspection revealed that some questions in our dataset were not genuinely related to robotics. For instance, this question \href{https://stackoverflow.com/questions/7861132}{\(Q_{7861132}\)}\footnote{https://stackoverflow.com/questions/7861132} had nothing to do with robotics despite having the \emph{robotics} tag. This question was asking about traversing nodes of a graph, and apart from the \emph{robotics} tag it does not mention a robot at all. Therefore, any conclusions drawn from this dataset might be inaccurate. To address this issue, we created a new, refined list of 500 verified robotics questions. 

To create this list, one of the authors, a senior PhD student in robotics with extensive experience in advanced software engineering, randomly selected 300 questions from the original 1,402. Each sampled item was subsequently examined manually, and questions deemed unrelated to robotics were excluded. This screening process yielded a corpus of 224 robotics-relevant questions and required approximately 15 hours to complete. Following this, the first and third authors applied the developed guidelines and worked together to expand the dataset by analyzing 365 additional questions. This yielded a final set of 500 validated robotics questions and required approximately 10 hours. These two authors stopped labeling more questions because this process was both time-consuming and labor-intensive. However, it is worth noting that similar research has successfully used even smaller datasets~\cite{Pinto:2014}.  Additionally, manual labeling of these 500 questions was necessary for addressing RQ2 and RQ4 (described later). Expanding the dataset would have required proportionally more manual effort, which was a limiting factor in our study.

%% file: 5_results.tex
\section{Analysis and Results}
\label{sec:results}

In this section, we present our approach to answering each research question along with the results. 


\begin{figure*}[h!]
    \centering
    \includegraphics[width=1\textwidth,keepaspectratio]{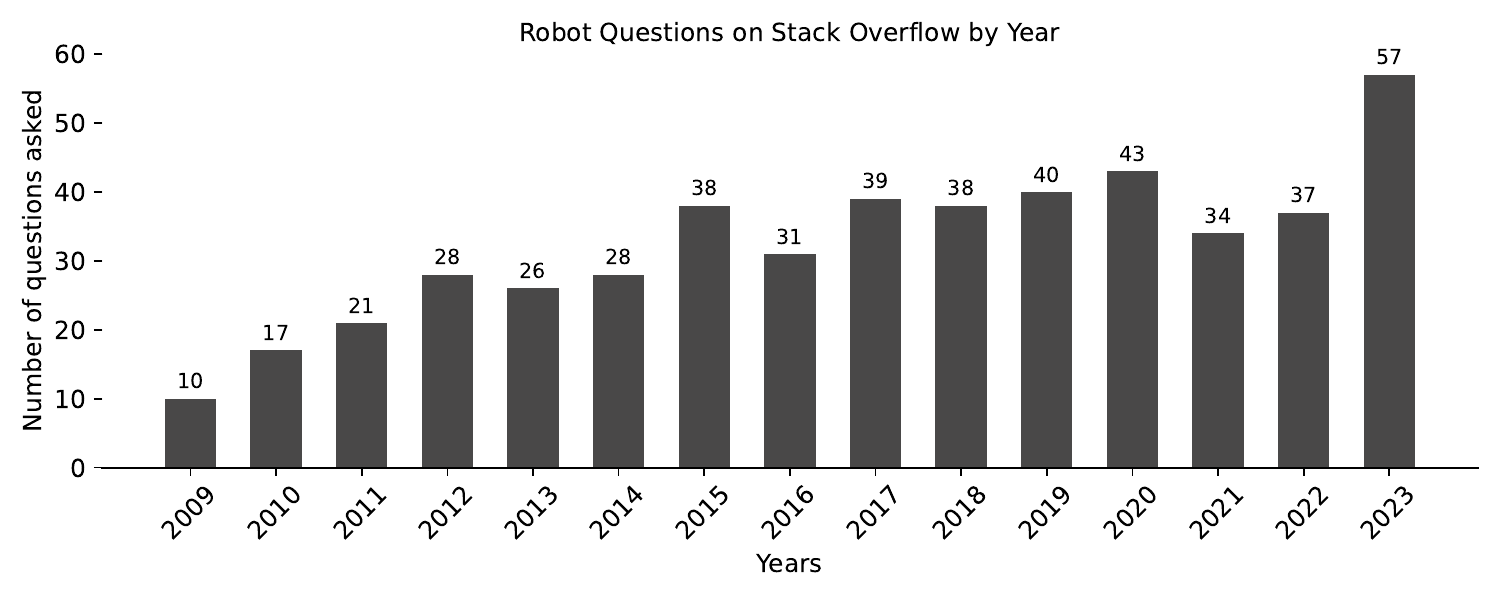}
    \caption{Number of robotics questions per year in the dataset of 500 selected questions. We excluded the 2024 data (13 questions) because our data collection ended in March 2024, and the partial year could bias the trend.}
    \label{fig:all_robot_q_per_yr_graph}    
\end{figure*}

\subsection{RQ1: Characteristics of the robotics questions}
\label{subsec:rq1}
The community consensus is that challenges in robotics development are distinct from those in other fields~\cite{goues2024software}. We are, therefore, interested to learn if these differences are reflected in how the robotics community asks questions and interacts with each other, and if these patterns are distinctive in nature when compared to other communities.

\subsubsection{Approach}
We first captured the posted date of the robotics questions to analyze whether interest in robotics questions is increasing or decreasing. Next, we compared the popularity of robotic questions to randomly selected one million \textsc{StackOverflow} questions. Encouraged by Pinto \textit{et al.}~\cite{Pinto:2014}, we used four popularity factors to calculate the overall popularity: \textit{score}, number of \textit{answers}, number of \textit{comments}, and number of \textit{views}. Pinto \textit{et al.}~\cite{Pinto:2014} also used the \textit{favourites} metric which is no longer available on \textsc{StackOverflow}. A user can \textit{up-vote} or \textit{down-vote} a question if they think the question is good or bad, respectively. The \textit{score} is the sum of these votes, with an \textit{up-vote} contributing \(+1\) and a \textit{down-vote} contributing \(-1\) to the score.

To facilitate a meaningful comparison with other \textsc{\textsc{StackOverflow}} questions, we focused on the relative popularity of the robotics questions which was calculated with the following equation : 

\begin{equation}\label{eqn:pinto_popularity}
    \mathbb{P} = \frac{\mathbb{S} + \mathbb{A} + \mathbb{C} + \mathbb{V}}{4}
\end{equation}
Here, \(\mathbb{S}\), \(\mathbb{A}\), \(\mathbb{C}\), and \(\mathbb{V}\) are the relative \textit{score}, relative number of \textit{answers}, relative number of 
 \textit{comments}, and relative number of \textit{views}, respectively. A relative value was calculated by the following equation:

\begin{equation}\label{eqn:normalization}
    relativeValue = avgRobot / avgSO
\end{equation}

Here, \textit{avgRobot} is the average of all values for a given factor (e.g., \textit{score}) in the robotics question set, whereas \textit{avgSO} is the average of all values for that factor in the non-robotics question set. For example, if the average views are 100 and 200 across our robotics and non-robotics \textsc{StackOverflow} dataset, respectively, then the relative number of \textit{views} is $100/200=0.5$. After calculating these values for all four factors, the final relative popularity is calculated with equation \ref{eqn:pinto_popularity}. This way, a relative popularity value of 0.5 indicates that robotics questions are half as popular as the average \textsc{StackOverflow} question, while a value of 2 indicates they are twice as popular.

Using the same approach, we calculate the relative popularity of answers for the robotics questions. As answers on \textsc{StackOverflow} do not have a \textit{view} count or \textit{answer} count, inspired by Pinto et al. \cite{Pinto:2014}, we converted the equation to the following:
\begin{equation}\label{eqn:answer_popularity}
    \mathbb{P} = \frac{\mathbb{S} + \mathbb{C}}{2}
\end{equation}

\subsubsection{Results}
Figure \ref{fig:all_robot_q_per_yr_graph} shows a visual representation of whether robotic questions are increasing or decreasing. This figure shows that, compared to the previous years, the number of robotics questions has increased significantly since 2015.  The data for 2023 shows a notable rise in the volume of such questions. We excluded the 2024 data (13 questions) from this figure because our data collection ended in March 2024, and the partial year could bias the trend. To estimate the full year's data, we can extrapolate by multiplying the 2024 count by 4, resulting in $13 \times 4 = 52$ questions. Overall, the evidence suggests an upward trend in the number of robotics questions on \textsc{StackOverflow}.

Despite the increasing number of robotics questions, their overall popularity remains lower compared to other \textsc{StackOverflow} questions, with a normalized popularity score of 0.67 (Table \ref{tab:all_pop_q_table}). This means that, on average, robotics questions are about \(\frac{2}{3}\) as popular as the typical \textsc{StackOverflow} question. This reflects the niche size of the robotics community and should not be interpreted as lower importance. Consistent with this, the \textit{score} and \textit{view} metrics for robotics questions are both relatively low, at 0.51, whereas the \textit{answer} and \textit{comment} metrics are higher, with values of 0.84 and 0.81, respectively. While robotics questions receive fewer up-votes and views, they tend to attract a relatively high number of answers and comments, even though these metrics are still lower than those for other types of questions. These findings suggest that, although the robotics community is small, its members are actively engaged in answering and commenting on robotics-related queries, indicating a dedicated and interactive community.

\begin{table}[t]
    \caption{Relative popularity and popularity factors of robotics questions.}
    \label{tab:all_pop_q_table}
    \begin{tabular}{ccccc}
        \toprule
        \textbf{Score} & \textbf{Answer} & \textbf{Comment} & \textbf{View} & \textbf{Popularity} \\
        \midrule
        0.51 & 0.84 & 0.81 & 0.51 & 0.67 \\
        \bottomrule
    \end{tabular}
\end{table}

\begin{table}[t]
    \caption{Relative popularity and popularity factors of the answers for the robotics questions.}
    \label{tab:all_pop_a_table}
    \begin{tabular}{ccc}
        \toprule
        \textbf{Score} & \textbf{Comment} & \textbf{Popularity} \\
        \midrule
        0.32 & 0.89 & 0.61 \\
        \bottomrule
    \end{tabular}
\end{table}



A similar trend is observed in the popularity factors for robotics answers, as detailed in Table \ref{tab:all_pop_a_table}.  Overall, robotics answers tend to be less popular than the average \textsc{StackOverflow} answer. Notably, the two metrics used to assess the popularity show a significant contrast for robotics: the \textit{score} is 0.32, while the \textit{comment} metric is 0.89. This disparity suggests that answering robotics questions may be more challenging, leading to lower scores. As a result, users are more inclined to engage by leaving comments—since comments are not subject to upvotes or downvotes, they do not affect reputation. Taken together, the low popularity metrics primarily reflect a niche domain rather than low relevance, and the comparatively high interaction levels highlight an active and supportive robotics community on \textsc{StackOverflow}.

\begin{tcolorbox}[colback=white, colframe=black, sharp corners]
\textbf{RQ1:} Robotics questions are increasing over time. However, robotic questions and answers are less popular than others. Although robotics questions are viewed less, they have relatively higher engagement in terms of answers and comments, indicating an engaged niche community. 
\end{tcolorbox}

\subsection{RQ2: Themes in Robotic Discussions}
\label{subsec:rq2}
Our RQ1 results show a distinct engagement pattern in how they are approached by the community. However, the specific types of questions that robotics developers ask are not yet fully understood. Gaining insight into these questions could illuminate the challenges developers face, enabling educators to tailor their curricula to address these issues more effectively. Additionally, these findings can help researchers align their studies with the most pressing concerns identified by developers.

\subsubsection{Approach}
\label{subsubsec:rq2_approach}
To identify the most common types of questions asked about robot development, we conducted a thematic analysis. Thematic analysis is used to identify patterns in the data which then become categories/themes to be used for analysis \cite{Fereday:2006}. Thematic analysis consists of six steps: familiarizing with the data, generating initial codes, searching for themes among codes, reviewing themes, defining and naming themes, and producing the final report \cite{Pinto:2014}. We created the themes inductively, that is, we created new themes only when a new question did not match any of the previous themes. The initial themes that were similar to each other were then grouped together to form larger \textit{themes} which were later labeled.

While forming the larger themes, we set a minimum threshold of 8 occurrences for a theme to be included in our analysis. This threshold was chosen based on a previous study by Pinto \textit{et al.}~\cite{Pinto:2014} that employed thematic analysis and used a threshold of 5 occurrences. That study analyzed 325 questions with labeled themes, while our analysis involved 500 questions. Therefore, we adjusted the threshold proportionally, calculating \(\frac{500}{325} \times 5 = 7.7 \approx \lceil 7.7 \rceil = 8\). This approach ensures that the themes identified are significant and not merely the result of random or infrequent occurrences. Themes that had fewer than 8 questions were merged together and labeled as \emph{other}.

To ensure the labeling of the themes was accurate and consistent, the second author, who is a senior PhD student in robotics, coded the first 224 questions and created sub-themes and themes from them, a process that required approximately 20 hours. During this initial coding, semantic ambiguity and terminological variation were addressed by consolidating synonymous expressions under shared concept definitions and maintaining a brief codebook that documented inclusion criteria and representative examples for each theme. The first and third authors, two undergraduate research assistants in software engineering, then familiarized themselves with these 224 questions, the resulting thematic framework, and the associated codebook, with support from the last author. Subsequently, the first and third authors jointly discussed and coded the remaining 276 questions, applying the same concept-level mapping to ensure that different surface forms (e.g., alternative technical terms or synonyms) were coded consistently. During this phase, they encountered ten instances of disagreement, which were addressed through structured meetings in which each question was reviewed individually until consensus was achieved; the last author was consulted only in the rare cases where agreement was difficult to reach. This phase required approximately 10 hours per coder.


Inspired by Pinto \textit{et al.}~\cite{Pinto:2014}, we expanded upon the metrics used in RQ1 by introducing an additional metric to further analyze the various themes in robotics discussions. On \textsc{StackOverflow}, a user can mark an answer to their question as \textit{accepted}. This indicates that the user finds that specific answer helpful enough to resolve their issue. There can be at most one accepted answer per question. Using this, we define three success statuses of questions. A \textit{successful} question has an accepted answer; an \textit{ordinary} question has at least one answer but none of them are accepted; and an \textit{unsuccessful} question has no answer.


\subsubsection{Results}

\begin{figure*}[htbp]
    \centering
       \includegraphics[width=0.78\textwidth,keepaspectratio]{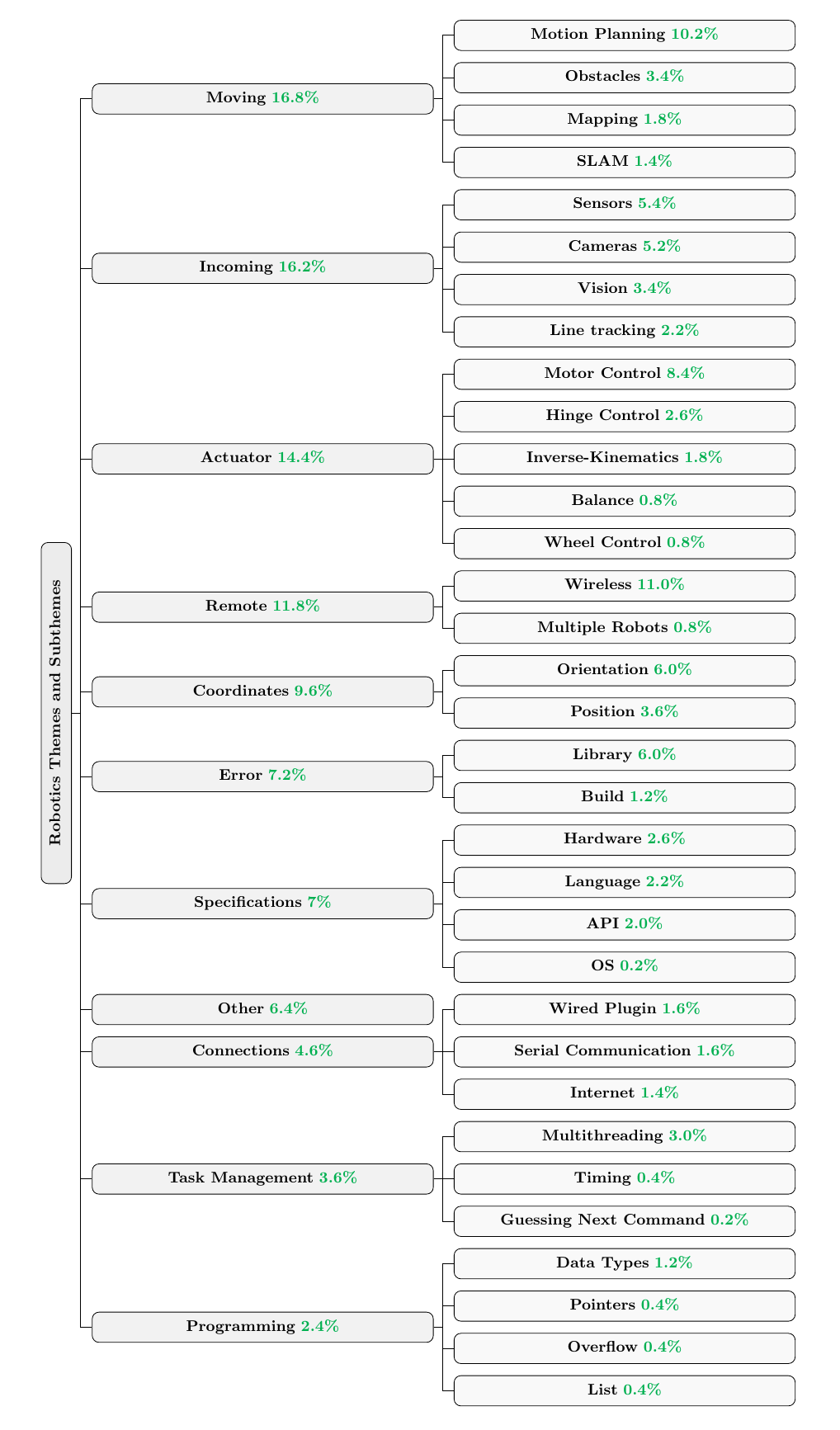}
       \caption{The distribution of all the 500 labeled questions based on the themes and sub-themes.}
    \label{fig:hierarchical_plot}
\end{figure*}

\begin{table*}[t]
\begin{center}
    \caption{Different popularity metrics for the 11 major themes. \textbf{Bolded} numbers indicate the highest value for each column while \textit{italicized} numbers indicate the lowest value for each column. \textbf{Q} is the number of questions, and \textbf{A} is the number of answers. $\mathbb{S}$, $\mathbb{A}$, $\mathbb{C}$, $\mathbb{V}$, and $\mathbb{P}$ are the same as in \textit{RQ1}. }
    \label{tab:themes_metrics_table}
    
        \begin{tabular}{l|ccc|ccc|cccc|c}
        \toprule
        \textbf{Themes} & \textbf{Q} & \textbf{A} & \textbf{A/Q} & \textbf{Unsuccessful} & \textbf{Ordinary} & \textbf{Successful} & $\mathbb{S}$ & $\mathbb{A}$ & $\mathbb{C}$ &  $\mathbb{V}$ & $\mathbb{P}$ \\
        \midrule

         Moving & \textbf{84} & \textbf{91} & 1.08 & 25.00\% & 36.90\% & 38.10\% & 0.49 & 0.73 & 0.78 & 0.30 & 0.58 \\
         Incoming & 81 & 90 & 1.11 & 24.69\% & 37.04\% & 38.27\% & 0.32 & 0.75 & 0.77 & 0.44 & 0.57 \\
         Actuator & 72 & 81 & 1.13 & 22.22\% & 47.22\% & 30.56\% & 0.30 & 0.76 & 1.00 & 0.24 & 0.58 \\
         Remote & 59 & 79 & 1.34 & 23.73\% & 45.76\% & \textit{30.51\%} & \textit{0.18} & 0.91 & 0.74 & 0.39 & 0.55 \\
         Coordinates & 48 & 47 & 0.98 & 29.17\% & 31.25\% & 39.58\% & 0.40 & 0.66 & 0.66 & 0.71 & 0.61 \\
         Errors & 36 & 33 & \textit{0.92} & \textbf{36.11\%} & 33.33\% & 30.56\% & 0.28 & \textit{0.62} & \textit{0.54} & 0.35 & \textit{0.45} \\
        Specifications & 35 & 84 & \textbf{2.40} & 22.86\% & 34.29\% & 42.86\% & \textbf{2.00} & \textbf{1.62} & 0.70 & 1.22 & \textbf{1.39} \\
        Other & 32 & 47 & 1.47 & 25.00\% & 37.50\% & 37.50\% & 1.05 & 0.97 & 0.85 & \textbf{1.22} & 1.02 \\
        Connections & 23 & 31 & 1.35 & 30.43\% & 34.78\% & 34.78\% & 0.54 & 0.91 & 1.00 & 0.77 & 0.81 \\
        Task Management & 18 & 22 & 1.22 & 27.78\% & \textit{16.67\%} & \textbf{55.56\%} & 0.42 & 0.83 & 0.95 & 0.33 & 0.63 \\
        Programming & \textit{12} & \textit{17} & 1.42 & \textit{8.33}\% & \textbf{50.00\%} & 41.67\% & 0.33 & 0.96 & \textbf{1.59} & \textit{0.13} & 0.75 \\
        \bottomrule
    \end{tabular}
    \end{center}
\end{table*}

Figure~\ref{fig:hierarchical_plot} shows the 11 major themes and 33 sub-themes. Themes and sub-themes are ordered in descending order from the top. Next, we discuss the themes and provide some example questions for each theme. 

\textbf{Moving (16.8\%):} 
These questions were related to a robot's movement, \example{... if I want to repeat the movement set going from A to D only, how can I do this using the information I have already gathered. ...} and were the most common. \textit{Motion Planning (10.2\%)} questions were the most common in the theme \textit{Moving}. These questions were about making a robot find the shortest path \href{https://stackoverflow.com/questions/742369}({\(Q_{742369}\)}\footnote{https://stackoverflow.com/questions/742369/}), finding a flag (\href{https://stackoverflow.com/questions/5361791}{\(Q_{5361791}\)}), and other general problems with the robot navigating through paths (\href{https://stackoverflow.com/questions/12994888}{\(Q_{12994888}\)}). \textit{Obstacles} contains questions about detecting and avoiding obstacles (\href{https://stackoverflow.com/questions/6641055}{\(Q_{6641055}\)},  \href{https://stackoverflow.com/questions/13677658}{\(Q_{13677658}\)}). 
Another subtheme was \textit{Mapping} containing questions related to how to map a room (\href{https://stackoverflow.com/questions/54652810}{\(Q_{54652810}\)}) or creating maps or occupancy grids (\href{https://stackoverflow.com/questions/58056874}{\(Q_{58056874}\)}). The last subtheme \textit{SLAM} contains questions that are very specific to \emph{Simultaneous localization and mapping}\footnote{https://www.mathworks.com/discovery/slam.html (last accessed: 23-Aug-2024)} (\href{https://stackoverflow.com/questions/40584193}{\(Q_{40584193}\)},  \href{https://stackoverflow.com/questions/74133720}{\(Q_{74133720}\)}).

\textbf{Incoming (16.2\%):}
These questions are related to receiving and using incoming data, \example{I want to calibrate the camera and find the transformation from camera to end-effector.}. The data could come from \textit{Cameras} (\href{https://stackoverflow.com/questions/62046666}{\(Q_{62046666}\)}, \href{https://stackoverflow.com/questions/67072289}{\(Q_{67072289}\)}) or \textit{Sensors} (\href{https://stackoverflow.com/questions/6620778}{\(Q_{6620778}\)},  \href{https://stackoverflow.com/questions/10075285}{\(Q_{10075285}\)}) on the robot. If a question was about software or some tracking systems using the cameras, it was labeled as \textit{Vision} (\href{https://stackoverflow.com/questions/14041109}{\(Q_{14041109}\)}, \href{https://stackoverflow.com/questions/21059100}{\(Q_{21059100}\)}). In cases where a robot was to follow lines on the ground, they were labeled as \textit{Line Tracking} (\href{https://stackoverflow.com/questions/54242283}{\(Q_{54242283}\)},  \href{https://stackoverflow.com/questions/1590073}{\(Q_{1590073}\)}).

\textbf{Actuator (14.4\%):}
These questions are about controlling parts of a robot that moved, \example{... What I want is that: The robot should move forward until it detects the object. When object is detected then it should stop the the wheels and start the arm to lift the object upward. ...}. Questions that were about any general issues with the motors of a robot were termed \textit{Motor Control} (\href{https://stackoverflow.com/questions/18718014}{\(Q_{18718014}\)}, \href{https://stackoverflow.com/questions/57038494} {\(Q_{57038494}\)}). Questions that required the usage of \textit{Inverse-Kinematics} (\href{https://stackoverflow.com/questions/2625390}{\(Q_{2625390}\)}, \href{https://stackoverflow.com/questions/53493976}{\(Q_{53493976}\)}) equations were labeled as such. If the question was more specific about \textit{Hinge Control} (\href{https://stackoverflow.com/questions/10420966}{\(Q_{10420966}\)}, \href{https://stackoverflow.com/questions/74063852}{\(Q_{74063852}\)}), \textit{Balance} (\href{https://stackoverflow.com/questions/16664330}{\(Q_{16664330}\)}, \href{https://stackoverflow.com/questions/76226178}{\(Q_{76226178}\)}), or \textit{Wheel Control} ( \href{https://stackoverflow.com/questions/6619222}{\(Q_{6619222}\)},  \href{https://stackoverflow.com/questions/3621244}{\(Q_{3621244}\)}), they were termed as such.

\textbf{Remote (11.8\%):}
These questions involved controlling a robot using various methods, \example{NXT bluetooth pairing always failed}. These methods include wired or wireless (\href{https://stackoverflow.com/questions/22003790}{\(Q_{22003790}\)}, \href{https://stackoverflow.com/questions/35374892}{\(Q_{35374892}\)}) methods such as a computer, phone, remote control, or joystick. If many robots were to be controlled, they were termed as \textit{Multiple Robots} (\href{https://stackoverflow.com/questions/29399433}{\(Q_{29399433}\)}).

\textbf{Coordinates (9.6\%):}
These questions were about the coordinates of a robot, \example{I am trying to implement a basic path planning algorithm . . . The only thing required for my algorithm is to make my robot oriented towards a given (using mouse) point in the x,y plane.}. This included determining a robot's \textit{Orientation} (\href{https://stackoverflow.com/questions/60682233}{\(Q_{60682233}\)}) or its \textit{Position} (\href{https://stackoverflow.com/questions/44097582}{\(Q_{44097582}\)}, \href{https://stackoverflow.com/questions/74667949}{\(Q_{74667949}\)}).



\textbf{Errors (7.2\%):}
These questions are about fixing errors or failures developers encountered when using a third-party (\href{https://stackoverflow.com/questions/29615931}{\(Q_{29615931}\)}, \href{https://stackoverflow.com/questions/37649397}{\(Q_{37649397}\)}) library, \example{... This is the command that I run: ... And this the error that I have: ...}.

\textbf{Specifications (7\%):} 
This theme contains questions about what to use when building a robot, \example{What is the right RTOS for a humanoid robot?}. This included getting \textit{Hardware} specifications (\href{https://stackoverflow.com/questions/9614729}{\(Q_{9614729}\)}), library or \textit{API} selection (\href{https://stackoverflow.com/questions/69371534}{\(Q_{69371534}\)}, \href{https://stackoverflow.com/questions/78227765}{\(Q_{78227765}\)}), \textit{Language} selection (\href{https://stackoverflow.com/questions/7017136}{\(Q_{7017136}\)}, \href{https://stackoverflow.com/questions/12462461}{\(Q_{12462461}\)}), or \textit{OS} recommendations (\href{https://stackoverflow.com/questions/7804388}{\(Q_{7804388}\)}) for building robots.

\textbf{Other (6.4\%):}
These questions did not occur enough to be considered their own separate theme. For example, \href{https://stackoverflow.com/questions/64299715}{\(Q_{64299715}\)} talks about optimization, \example{Is there a way to simplify repetitive addition in x-drive?}. Details on these sub-themes' can be found in our shared public repository.

\textbf{Connections (4.6\%):}
This theme includes questions related to challenges in connecting a robot to  a network, \example{... I tried to stream the video from the camera using the http live streaming protocol and vlc, but the latency is too high (15-30sec) to properly control it. ...}. Sub-themes in this theme are, \textit{Internet} (\href{https://stackoverflow.com/questions/6140332}{\(Q_{6140332}\)}, \href{https://stackoverflow.com/questions/58836477}{\(Q_{58836477}\)}), connecting using a \textit{Wired Plugin} method (\href{https://stackoverflow.com/questions/62414292}{\(Q_{62414292}\)}, \href{https://stackoverflow.com/questions/78212426}{\(Q_{78212426}\)}), and \textit{Serial Communication} (\href{https://stackoverflow.com/questions/60640639}{\(Q_{60640639}\)}) (e.g., uploading code to the robot).

\textbf{Task Management (3.6\%):}
These questions concern the coordination of multiple tasks or motions within a robotics system, \example{I have two C++ classes, objecManip and updater. The updater class has a timer to check the status of the robot arm of my application. ... The problem is that when filling in actions queue current\_status is taken constantly not dynamically.}.
This theme includes coordinating the
\textit{Timing} (\href{https://stackoverflow.com/questions/13856211}{\(Q_{13856211}\)},  \href{https://stackoverflow.com/questions/15320030}{\(Q_{15320030}\)}) of tasks such as the movement of the robot while also processing input. These questions also include wanting to know how to incorporate \textit{Multithreading} (\href{https://stackoverflow.com/questions/786997}{\(Q_{786997}\)}, \href{https://stackoverflow.com/questions/27716172}{\(Q_{27716172}\)}) into robot program or getting a robot to \textit{Guess Next Command} (\href{https://stackoverflow.com/questions/72090152}{\(Q_{72090152}\)}).

\textbf{Programming (2.4\%):}
These are questions related to programming, \example{I'm studying robotics at the university and I have to implement on my own SLAM algorithm. To do it I will use ROS, Gazebo and C++. I have a doubt about what data structure I have to use to store the map.}. These occur in other areas outside of robotics as well, such as questions about \textit{Pointers} (\href{https://stackoverflow.com/questions/16419356}{\(Q_{16419356}\)},  \href{https://stackoverflow.com/questions/64457784}{\(Q_{64457784}\)}), \textit{Data Types} (\href{https://stackoverflow.com/questions/57675404}{\(Q_{57675404}\)},  \href{https://stackoverflow.com/questions/64552488}{\(Q_{64552488}\)}), \textit{Overflow} (\href{https://stackoverflow.com/questions/21999921}{\(Q_{21999921}\)}, \href{https://stackoverflow.com/questions/65099688}{\(Q_{65099688}\)}), and \textit{List} (\href{https://stackoverflow.com/questions/48227848}{\(Q_{48227848}\)},  \href{https://stackoverflow.com/questions/66120048}{\(Q_{66120048}\)}). It is worth noting that just because a question has code contained in it, it does not become a \textit{Programming} question. For instance, \href{https://stackoverflow.com/questions/18718014}{\(Q_{18718014}\)} has code in it but it is not a programming question as the main problem is controlling the motor of the robot. Additionally, had we included all questions that had code as the \textit{Programming} category, too many questions would fall into that one category, skewing the results and making them less useful.

We now discuss different popularity metrics of the 11 major themes, as presented in Table~\ref{tab:themes_metrics_table}. Clearly, the numbers of questions in these themes are not evenly distributed. \textit{Moving} has the highest number of questions (84), probably because it is one of the fundamental aspects of robotics, essential for many applications in the field~\cite{niku2020introduction}. Usually, robotic practitioners build robots that involve moving through space, such as navigating through a maze. The \textit{Actuator} theme is related to the moving parts of a robot, such as motors and joints, while the \textit{Incoming} theme deals with data from cameras and sensors, which is often essential for navigation and movement. The fact that these top three themes have similar question numbers underscores the community's strong interest in the movement aspects of robotics. And their answer-to-question ratio (A/Q) and other popularity metrics reveal the challenges that robotics developers face.

In contrast, the \textit{Programming} theme has the lowest number of questions (12) but boasts a high percentage of both ordinary and successful inquiries. This suggests that practitioners face fewer challenges in this area, and when they do post questions, they are frequently answered. This is intuitive because these questions can usually be answered using the plethora of resources available, including the millions of questions about coding on \textsc{StackOverflow}. Due to this, there is usually not much need to ask about \textit{Programming} questions separately with the robot or robotics tags. Surprisingly, although the \textit{Errors} theme is the closest to the \textit{Programming} theme, it has the highest percentage of unsuccessful questions (36.11\%) and the lowest A/Q ratio (0.92). This implies that robotics-specific libraries and build error-related questions are often complex making them difficult to answer.



\textit{Task Management} has the highest percentage of accepted answers suggesting that, when questions are asked about this topic, they are resolved effectively, indicating clear solutions. Also, since there is a low percentage of ordinary questions, that suggests that questions in this category are often polarized, they either have a really good answer (accepted) or have no answers. 

\textit{Specifications} having the highest score in three of the four popularity metrics indicates that it is an overall popular topic, and its popularity is not falsely inflated by certain metrics. \textit{Specifications} has a high A/Q ratio (2.40), indicating there is a high level of engagement as well as many people being knowledgeable in this topic, thus contributing to a high A/Q ratio. This is probably one of the reasons behind practitioners asking fewer questions about \textit{Specifications} in recent times---will show in RQ3.

We also analyzed the popularity of the answers to the robotics questions (Table~\ref{tab:themes_a_popularity_table}). Answers to the \textit{Programming} questions, despite their low scores, are the most popular, whereas answers to the \textit{Errors} questions are the least popular. If we consider the scores only, the difference between up-votes and down-votes, answers to the \textit{Specifications} questions are the most popular.


\begin{table}[htbp]
\centering
    \caption{Popularity metrics of the answers for the identified themes (ranked with the overall popularity).}
    \label{tab:themes_a_popularity_table}

    \begin{tabular}{lccc}
        \toprule
        \textbf{Themes} & $\mathbb{S}$ & $\mathbb{C}$ & $\mathbb{P}$ \\
        \midrule
        Programming & \textit{0.09} & \textbf{2.52} & \textbf{1.30} \\
        Task Management & 0.43 & 1.68 & 1.06 \\
        Specifications & \textbf{0.95} & 0.96 & 0.96 \\
        Connection & 0.18 & 1.15 & 0.67 \\
        Other & 0.64 & 0.59 & 0.62 \\
        Incoming & 0.28 & 0.94 & 0.61 \\
        Moving & 0.32 & 0.88 & 0.60 \\
        Coordinates & 0.26 & 0.87 & 0.57 \\
        Remote & 0.17 & 0.82 & 0.49 \\
        Actuator & 0.20 & 0.73 & 0.47 \\
        Errors & 0.24 & \textit{0.26} & \textit{0.25} \\
        \bottomrule
        
    \end{tabular}
\end{table}

\begin{tcolorbox}[colback=white, colframe=black, sharp corners]
\textbf{RQ2:}  We found the theme \textit{Moving (16.8\%)} had the most occurrences of questions and the subtheme with the most questions was \textit{Wireless (11.0\%)}. In general, robotics practitioners frequently face challenges with \textit{Moving}, \textit{Actuator}, and \textit{Incoming} related questions. \textit{Specifications} and \textit{Programming} questions are, in general, easy to answer.
\end{tcolorbox}


\subsection{RQ3: Evolution of the identified themes}
\label{subsec:rq3}

Inspired by similar studies~\cite{Han:2020, Uddin:2021, Barua:2014}, we examine the time-varying impact of themes in robotics-related questions to move beyond overall frequencies and capture how practitioner concerns shift over time. This analysis helps identify persistent pain points, assess the influence of emerging technologies on developer challenges, and detect declining themes that may reflect tool and knowledge maturation. These temporal insights highlight where research and engineering effort is most needed and provide actionable guidance for educators and practitioners in prioritizing current competencies and learning needs.

\subsubsection{Approach}

To analyze trends and the time-varying impacts of various robotics themes, we examined the yearly probabilities for each theme individually. We also computed the relative probability of each theme in comparison to others. Equation~\ref{eqn:prob1} calculates the probability of a theme $t$ in a specific year $y_i$ based on the number of questions related to that theme posted during our data collection period (2009 to 2024). Here, $N_{ty_i}$ represents the number of questions on theme $t$ for year $y_i$. 

A graph derived from Equation~\ref{eqn:prob1}—the Probability Distribution Function (PDF)~\cite{ash2008basic}—will illustrate the yearly trends of various themes posted on \textsc{StackOverflow}. However, PDFs are not monotonic and often display a zigzag pattern, which can hinder readability. Therefore, we utilize the Cumulative Distribution Function (CDF)~\cite{deisenroth2020mathematics}, as outlined in Equation~\ref{eqn:prob2}. Here, $C_{ty}$ is the sum of probabilities of a theme $t$ from year 2009 to year $y$.
PDF and CDF plots are commonly used in prior studies, including software engineering studies and empirical studies of \textsc{StackOverflow} to characterize distributional properties in a visually stable way~\cite{asaduzzaman2013answering, chowdhury2024method, ahmed2025exploring, kavaler2018determinants, bhat2014min}.

Likewise, Equation~\ref{eqn:prob3} calculates the relative probability of a theme in relation to others. Here, $N_{ty_i}$ and $N_{Ty_i}$ are the numbers of questions posted in year $y_i$ on themes $t$ and all themes, respectively. It is important to note that we cannot create a CDF for this equation, as the CDF is only applicable when evaluating a theme in isolation.


\begin{equation}\label{eqn:prob1}
    P_{ty_i} = \frac{N_{ty_i}}{\sum_{y_i=2009}^{2024} N_{ty_i}}
\end{equation}

 \begin{equation}\label{eqn:prob2}
     C_{ty}  = \sum_{y_i=2009}^{y} P_{ty_i}
 \end{equation}

\begin{equation}\label{eqn:prob3}
    A_{ty_i}  = \frac{N_{ty_i}}{N_{Ty_i}}
\end{equation}

\subsubsection{Results}
\label{subsubsec:rq2_results}





\begin{figure*}[htbp]
\centering
\includegraphics[width=0.85\textwidth, keepaspectratio]{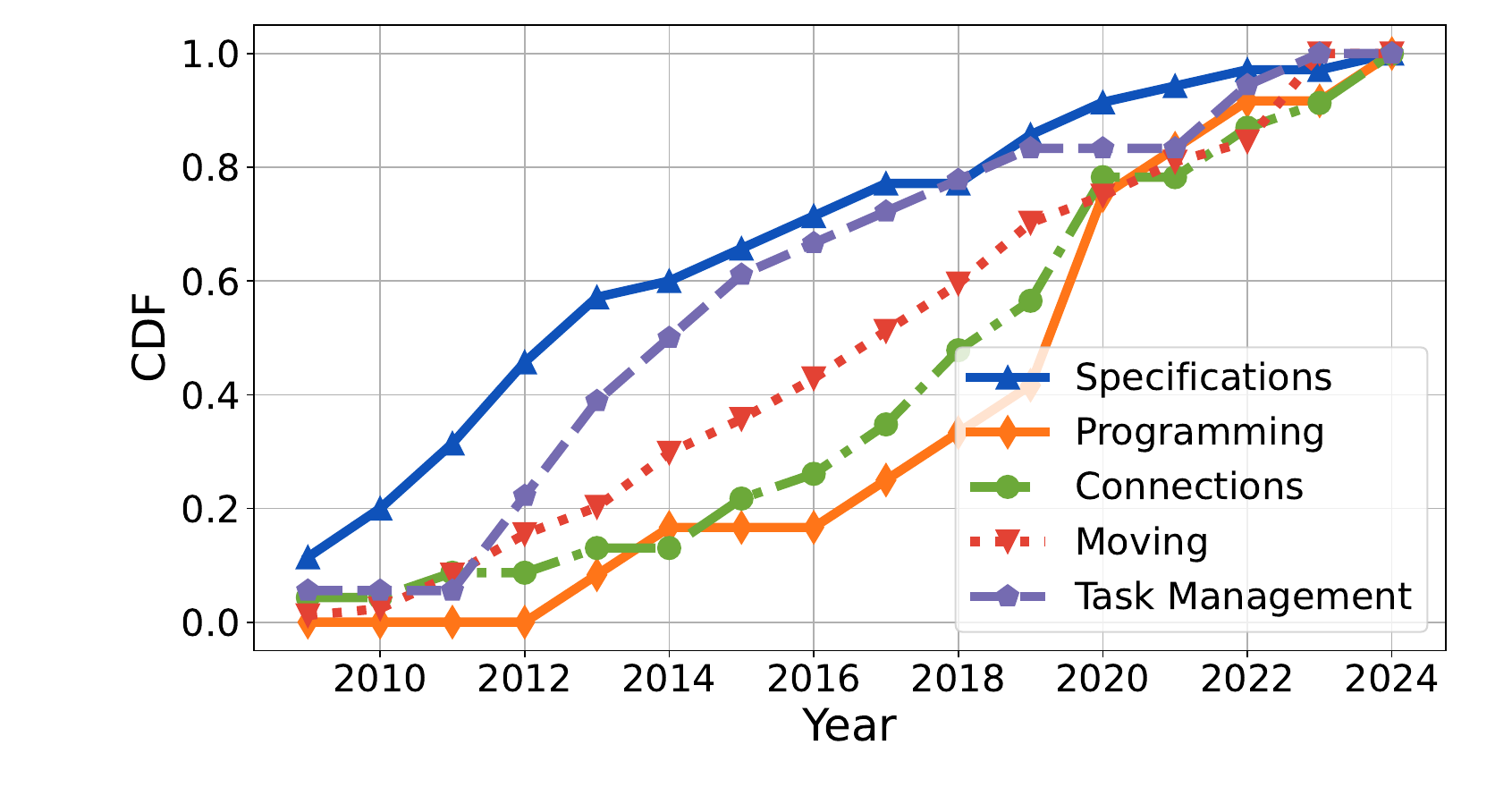}    
\caption{Cumulative Distribution Function (CDF) of the yearly probability of questions posted on different themes. To enhance graph readability, only five selected themes are presented, while other themes are discussed in the text. While some themes show skewed distributions, themes like \textit{Moving} exhibit stable popularity distribution throughout the years.}
\label{fig:cdf_evolution}
\end{figure*}

Figure~\ref{fig:cdf_evolution} shows the cumulative distribution function (CDF) of questions over time. For readability, we report only five representative themes. The x-axis denotes calendar year, and the y-axis indicates the cumulative percentage of questions posted up to and including each year.


Among the eleven themes, \textit{Specifications} and \textit{Task Management} exhibit distinct temporal patterns. Questions in these themes are concentrated in earlier years: by 2014, roughly 60\% of \textit{Specifications} questions and 50\% of \textit{Task Management} questions had been posted, whereas only about 20\% in each theme were posted after 2018. This suggests these topics were more salient earlier, with attention shifting to other themes in recent years. One explanation is that they reflect foundational concerns; consistent with this, \textit{Specifications} has the highest question popularity score (1.39), and \textit{Task Management} shows comparatively high engagement in answers (0.83) and comments (0.95) despite moderate question popularity (0.63) (Table~\ref{tab:themes_metrics_table}). Another explanation is that these issues are now well covered: \textit{Specifications} has the highest answer-to-question ratio (2.4) and a strong success rate (42.86\%), while \textit{Task Management} has the highest success rate (55.56\%); both also rank the top in answer popularity in Table~\ref{tab:themes_a_popularity_table} (1.30 and 1.06, respectively).

In contrast, questions associated with \textit{Programming} and \textit{Connections} are largely absent in the early years (prior to 2012) and then increase gradually over time. A comparable upward trajectory is observed for \textit{Error} and \textit{Coordinates}. Notably, \textit{Programming} shows a marked rise beginning around 2016, temporally aligning with the release of the PyTorch deep-learning framework~\footnote{https://github.com/pytorch/pytorch/releases?page=7}, followed by a sharper increase from 2019 onward, which coincides with the broader transition from ROS~1 to ROS~2 after the initial ROS~2 release in late 2017~\footnote{https://docs.ros.org/en/rolling/Releases.html}.

Questions concerning \textit{Moving} and the other four themes (i.e., \textit{Actuator}, \textit{Remote}, {Incoming}, and \textit{Other}) were distributed almost uniformly throughout the data collection period, suggesting that practitioners have consistently faced challenges in these areas from the beginning to the end of the study. 

\begin{figure*}[htbp]
\centering
\includegraphics[width=0.78\textwidth ,keepaspectratio]{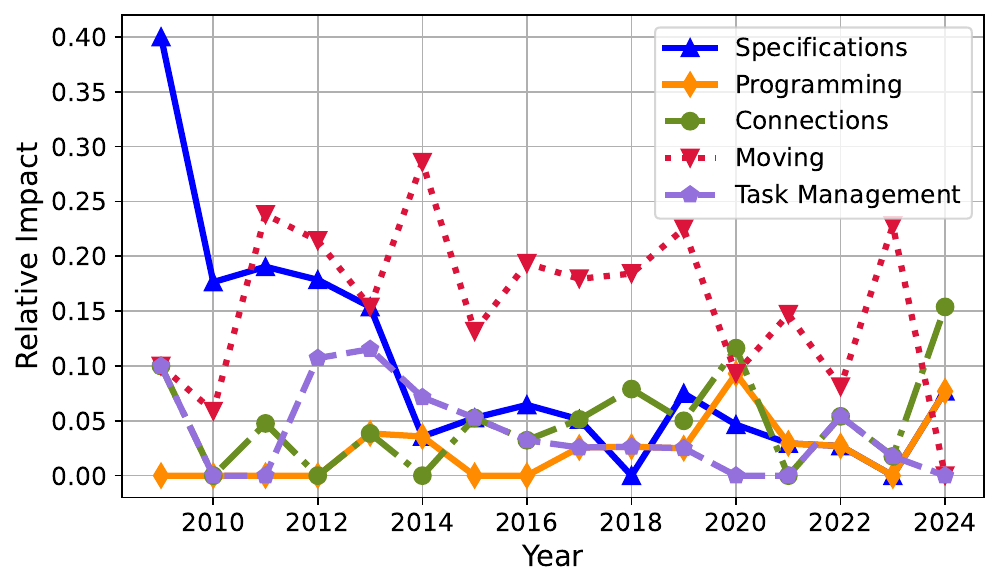} \caption{Relative impact per year for each theme. To improve graph readability, only a few themes are shown. \textit{Specifications} questions, despite their high early impact, lost traction after 2013. Questions related to robot \textit{Moving} always had their impact within the robotics community.}
\label{fig:rel_impact_graph}
\end{figure*}

Although the preceding analysis characterizes how themes evolved over time, it does not capture their time-varying prominence relative to one another. To address this limitation, we compute the relative impact of each theme—defined as the probability of observing a given theme in a particular year relative to all themes in that year—using Equation~\ref{eqn:prob3}. This metric identifies which themes were discussed most frequently each year, thereby complementing and strengthening the temporal trends observed earlier and highlighting topics that may represent persistent developer pain points warranting greater research attention. The results are shown in Figure~\ref{fig:rel_impact_graph}. For readability, the figure visualizes only a subset of themes; however, we discuss all themes here.

In 2009, most questions were associated with \textit{Specifications}. Although \textit{Specifications} remained a dominant theme through 2013, its relative impact declines markedly from 2014 onward, consistent with the CDF trends in Figure~\ref{fig:cdf_evolution}. In contrast, \textit{Moving}, \textit{Incoming}, \textit{Actuator}, and \textit{Remote} consistently exhibit high relative impact across the full study period, suggesting that these topics represent persistent developer pain points. Other themes continue to appear intermittently; however, they do not emerge as leading drivers of robotics-related discussions on \textsc{StackOverflow}.

To observe the impact in the most recent years, for each theme, we also analyzed the difference between the number of posted questions in the early years and the most recent years. The results are presented in Table~\ref{tab:diff_abs_impact}. In addition to their consistent impact, themes such as \textit{Actuator, Moving, Remote}, and \textit{Incoming} had significant impact in recent years as well. Although the overall impacts are less for themes such as \textit{Coordinates} and \textit{Error}, their recent impacts, however, are noticeable, as many of their questions are from very recent years. 

\begin{table}[htbp]
    \centering
    \caption{Most recent impact of each theme compared to their very early impact. Results for 2024 were not considered because of incompleteness in data collection.}
\resizebox{\columnwidth}{!}{
    \begin{tabular}{l|cc|c}
        \toprule
        \textbf{Theme} & \textbf{2009-2011} & \textbf{2021-2023} & \textbf{Difference} \\
        \midrule
        Actuator        & 4  & 22 & \textbf{18} \\
        Incoming        & 6  & \textbf{24} & \textbf{18} \\
        Moving          & 7  & 21 & 14 \\
        Error           & 0  & 13 & 13 \\
        Coordinates     & 2  & 14 & 12 \\
        Remote          & 7  & 18 & 11 \\
        Programming     & \textit{0}  & \textit{2}  & 2  \\
        Task Management          & 1  & 3  & 2  \\
        Connections     & 2  & 3  & 1  \\
        Other           & 8  & 6  & -2 \\
        Specifications  & \textbf{11} & \textit{2}  & \textit{-9} \\
        \midrule
        Mean            & 4.4 & 11.6 & 7.3\\
        Median          & 4 & 13 & 11\\
        \bottomrule
    \end{tabular}
    }
    \label{tab:diff_abs_impact}
\end{table}

\begin{tcolorbox}[colback=white, colframe=black, sharp corners]
\textbf{RQ3:} We found that the themes consistently dominating robotics discussions are \textit{Moving}, \textit{Actuator}, \textit{Incoming}, and \textit{Remote}. In recent years, questions related to \textit{Error} and \textit{Coordinates} have begun to gain traction. Although robotics practitioners initially faced significant challenges with \textit{Specifications} issues—such as Hardware, API, Languages, and OS—these concerns appear to have diminished since 2014.

\end{tcolorbox}

\subsection{RQ4: Types and Cognitive load in robotics questions}
\label{subsec:rq4}

Understanding the specific types of questions that practitioners in robotics ask is crucial for identifying knowledge gaps in this rapidly evolving field. Analyzing these question types—such as queries on implementation processes (\textit{How}), requests for information (\textit{What}), and inquiries into problem origins (\textit{Why})—is essential. This analysis goes beyond merely recognizing thematic patterns in our data; it provides educators with insights needed to develop targeted educational resources. For example, if there are a lot of questions of the type \textit{Why}, that could indicate educators should focus on addressing the reasoning behind what they do when making robots. If developers ask many \emph{what} questions, they need guidelines in selecting architectures, methods, and tool support or to understand requirements for their project completion. Also, by looking at the distributions of these types of questions, we can understand the required cognitive effort~\cite{paas1994instructional,van2005cognitive,bowell2005critical} to answer robotic questions: given that \emph{why} questions and \emph{what} questions require the highest and lowest cognitive efforts, respectively. 
By creating tailored tutorials, books, videos, and other materials, educators can effectively address the challenges faced by robotics developers, ultimately advancing learning and innovation in the field.

\subsubsection{Approach}


Inspired by multiple similar studies~\cite{Rosen:2016, Abdellatif:2020, Uddin:2021}, we categorize our 500 selected questions into four different categories: \textit{What}, \textit{How}, \textit{Why}, and \textit{Other}. The first and third authors were trained by the last author to apply the four-category scheme and developed operational definitions to guide consistent labeling. When a question expressed multiple or ambiguous intents, a label was assigned based on the primary information need reflected in the question. The first and third authors then collaboratively labeled all 500 questions and encountered 10 disagreements. These cases were resolved through a consensus-based process in which the coders compared rationales and consulted the operational definitions; the last author adjudicated the rare unresolved cases. Final labels were assigned once agreement was reached. This labeling process required approximately 15 hours. Table~\ref{tab:hwwo} presents the question-type taxonomy, along with its definitions and representative examples.

\input{5.1_table}

\textbf{What} questions aim to gather specific information about something, require help in a decision, request recommendations for a tool to build a robot, and inquire about the availability of a tool. Moreover, these questions tend to focus on theoretical concepts and abstract ideas. 

The most straightforward \textit{What} questions are those that have \emph{What} in the question sentence such as in the questions \href{https://stackoverflow.com/questions/40419901}{\(Q_{40419901}\)}, \href{https://stackoverflow.com/questions/41604310}{\(Q_{41604310}\)}, \href{https://stackoverflow.com/questions/4032394}{\(Q_{4032394}\)}. However, just because a user does not mention the word \emph{What} in their question does not mean it is not of this type. For example, the user in \href{https://stackoverflow.com/questions/45996174}{\(Q_{45996174}\)} presents multiple candidate communication methods and requests a recommendation regarding which option is more appropriate for robot–PC/Android communication; because the central intent is \emph{selecting among alternatives} rather than requesting step-by-step implementation guidance, we classify it as a \textit{What} question. Similarly, \href{https://stackoverflow.com/questions/47165908}{\(Q_{47165908}\)} is asking if there is a better alternative to their code. 



\begin{table}[t]
    \caption{Types of robotics questions asked by \textsc{StackOverflow} users.}
    \label{tab:HWWO_percentage}
\resizebox{\columnwidth}{!}{
\begin{tabular}{l|cccc}        
\toprule
        \textbf{Themes} & \textbf{\% How} & \textbf{\% What} & \textbf{\% Why} & \textbf{\% Other} \\
        \midrule
        Specifications       & \textit{12.12}    & \textbf{84.85}    & \textit{0.0}      & 3.03              \\
        Remote               & 56.90             & 34.48             & 3.45              & 5.17              \\
        Connections          & 39.13             & 26.09             & 13.04             & \textbf{21.74}    \\
        Coordinates          & \textbf{65.96}    & \textit{21.28}    & 6.38              & 6.38              \\
        Moving               & 53.57             & 39.29             & 5.95              & 1.19              \\
        Actuator             & 48.61             & 36.11             & 6.94              & 8.33              \\
        Programming          & 25.00             & 58.33             & 8.33              & 8.33              \\
        Task Management              & 50.00             & 33.33             & 5.56              & 11.11             \\
        Incoming             & 56.79             & 29.63             & 3.70              & 9.88              \\
        Errors               & 55.56             & 25.00             & 8.33              & 11.11             \\
        Other               & 41.67             & 44.44             & \textbf{13.89}    & \textit{0.00}     \\
        \midrule
        \textbf{Overall}     & \textbf{50.0}     & \textbf{37.0}     & \textbf{6.2}      & \textbf{6.8} \\
        \bottomrule
    \end{tabular}
    }
\end{table}

\textbf{How} questions are more direct and practical. They seek detailed explanations and step-by-step processes to achieve something. These inquiries are focused on understanding the precise methods required to achieve specific goals or outcomes. \href{https://stackoverflow.com/questions/54147301}{\(Q_{54147301}\)} and \href{https://stackoverflow.com/questions/38659146}{\(Q_{38659146}\)} are both examples of \textit{How} questions.


\textbf{Why} questions differ from other types of questions as they seek to understand the rationale behind a specific approach or the underlying causes of a problem. For example, \href{https://stackoverflow.com/questions/40584193}{\(Q_{40584193}\)} inquires about the reasoning behind calculating Jacobian matrices in EKF-SLAM. Likewise, \href{https://stackoverflow.com/questions/49842838}{\(Q_{49842838}\)} explores the question of why MPI is not a more commonly used approach.




The \textbf{Other} category is for questions that do not fall into any of the above categories. Most of the questions that resulted in this category was because the user does not specify what they are asking, thus leaving their post up to interpretation such as \href{https://stackoverflow.com/questions/15320030}{\(Q_{15320030}\)} and \href{https://stackoverflow.com/questions/16664330}{\(Q_{16664330}\)}. Sometimes, these questions can be seen as a mixture of the previous three types. For example, the exact question in \href{https://stackoverflow.com/questions/15320030}{\(Q_{15320030}\)}(“Real time programming in C++”) is vague; other users also indicated in the comments that additional details and clarification were necessary to understand the problem.

\subsubsection{Result} 

As shown in Table \ref{tab:HWWO_percentage}, our analysis reveals that 50\% of the questions asked by robotics practitioners on Stack Overflow are \textit{How} questions. This high proportion suggests that practitioners are primarily focused on obtaining practical, step-by-step guidance on various robotics topics. This emphasis on \textit{How} questions indicates a strong need for practical knowledge and hands-on solutions within the robotics community. This also means a large number of robotic questions require moderate cognitive effort to answer. 

Following this, 37\% of the questions are \textit{What} questions, suggesting a substantial demand for specific information, recommendations, and decision-making guidance, for example, on selecting one hardware tool over another.

The relatively small proportion of 31 \textit{Why} questions in total indicates that fewer users are concerned with understanding the underlying reason behind specific problems. Instead, there is a stronger emphasis on practical application and immediate solutions. 

A really interesting theme is the \textit{Specifications} theme. That theme consists of the highest percentage of \textit{What} questions (84.85\%) as well as the lowest percentage of \textit{How} (12.12\%) \textit{and} \textit{Why} (0.0\%) questions. This indicates a strong preference among users for seeking specific information or recommendations related to acquiring hardware, libraries, APIs, or details about hardware and operating systems. Notably, there are no questions asking \textit{why} should I use this specific tool or similar types of questions. Rather, this theme focuses solely on practical solutions in robotics.


The \textit{Coordinates} theme has the highest percentage of \textit{How} questions (65.96\%). Additionally, this theme also has the lowest percentage of \textit{What} questions (21.28\%). This indicates that many users who ask questions related to \textit{Coordinates} want to know step-by-step explanations in solving their problems rather than asking specific information on what the problem is. For instance, the user in \href{https://stackoverflow.com/questions/61409469}{\(Q_{61409469}\)} knows they want to make a contour mapping of a plot but they want assistance in \textit{how} to do it. Similarly, in \href{https://stackoverflow.com/questions/23009549}{\(Q_{23009549}\)}, the user knows to use certain formulas but they are unsure if those are the correct ones or not.



\begin{tcolorbox}[colback=white, colframe=black, sharp corners]
\textbf{RQ4:} Robot developers are mainly focused on learning hands-on implementation solutions by asking \textit{How} questions, followed by \textit{What} and \textit{Why}. This indicates that robotics practitioners need good detailed and accurate documentation, troubleshooting support, and support in requirement clarification. Encouragingly, it requires moderate cognitive effort to answer most of the robotics questions---as most of the robotics questions are in \text{How} category, not in \textit{Why} category. 
\end{tcolorbox}

%% file: 5.1_table.tex
\begin{table*}[ht]
    \centering
    \caption{Question-type taxonomy used in our analysis, summarizing the operational definitions for each category (What/How/Why/Other) and providing representative \textsc{StackOverflow} question IDs with an illustrative snippet.
}
    \label{tab:hwwo}
    \begin{tabular}{c|p{5cm}|c|p{5cm}}
\toprule
 Question
Type& Definitions& Sample Question&Example
Snippet\\
\midrule
         What&  Questions that request specific information, ask help in decision making, ask
definitions,  clarification or recommendation.&  \href{https://stackoverflow.com/questions/40419901}{\(Q_{40419901}\)} , \href{https://stackoverflow.com/questions/41604310}{\(Q_{41604310}\)} & What is difference of occupancy grid map and elevation map \\ \midrule
         How&  Questions that seek
detailed explanations and step-by-step processes to achieve
something.&  \href{https://stackoverflow.com/questions/54147301}{\(Q_{54147301}\)} ,  \href{https://stackoverflow.com/questions/38659146}{\(Q_{38659146}\)}& How to calculate the CTE with original co-ordinate (x1, y1, theta1) and current position (x1', y1', theta1')?\\ \midrule
         Why&  Questions that seek to understand the rationale behind a specific approach or the underlying causes of a problem. &  \href{https://stackoverflow.com/questions/40584193}{\(Q_{40584193}\)} , \href{https://stackoverflow.com/questions/49842838}{\(Q_{49842838}\)}& Why calculate jacobians in ekf-slam\\ \midrule
         Other&  Questions that do not fall
into any of the above categories and may involve open-ended discussions, opinions, or multiple unrelated questions. & \href{https://stackoverflow.com/questions/15320030}{\(Q_{15320030}\)} , \href{https://stackoverflow.com/questions/16664330}{\(Q_{16664330}\)} & Real time programming in C++ (Note: The exact question in this post is unclear)\\ \bottomrule
    \end{tabular}
\end{table*}

%% file: 6_discussion.tex
\section{Implications and Future Work}
\label{sec:discussion}
\revision{Robotics practitioners face significant challenges due to the limited advancements in software engineering specifically tailored for robot development~\cite{goues2024software}. In this mixed-methods study, we utilized both quantitative and qualitative research approaches to analyze the characteristics and themes of robotics-related questions posted on \textsc{StackOverflow}. Our findings reveal robotics questions are increasing over time but the questions and answers are less popular than average \textsc{StackOverflow} questions indicating a niche community (\textbf{RQ1}).  We identified that most robotics questions pertain to topics such as \textit{Moving}, \textit{Actuator}, \textit{Remote}, and \textit{Incoming} (\textbf{RQ2}). These topics have consistently posed challenges for practitioners over the years, including in recent times (\textbf{RQ3}). Additionally, our analysis shows that practitioners predominantly ask \textit{How} questions, compared to \textit{What} and \textit{Why} questions (\textbf{RQ4}). The findings have important implications for both research and practice.}  

\revision{For researchers, first, the increasing demand for robotics support on \textsc{StackOverflow} (\textbf{RQ1}) suggests continued growth, potentially accelerated by the convergence of robotics with contemporary AI methods. Second, robotics posts show low passive visibility but comparatively high active engagement (answers and comments), and answers attract substantial comment activity despite low scores (\textbf{RQ1}). This pattern indicates that view- and vote-based popularity is a weak proxy for practical value in niche domains; accordingly, researchers should incorporate interaction- and resolution-oriented measures to avoid systematically understating impact.}

\revision{Third, consistently elevated comment activity (\textbf{RQ1}, \textbf{RQ2}) suggests that robotics help-seeking frequently requires iterative clarification and additional contextual information. Studies that operationalize difficulty, resolvability, or knowledge-base quality should explicitly account for this back-and-forth dynamic.}

\revision{Fourth, support needs vary substantially by theme. Error-related posts exhibit weaker engagement and lower resolvability than other categories (\textbf{RQ2}), implying that failure diagnosis may be ill-suited to \textsc{StackOverflow}-style Q\&A without richer diagnostic artifacts. By contrast, \textit{Task Management} posts, though less frequent, are more likely to reach accepted solutions (\textbf{RQ2}). Collectively, these results argue for theme-specific interventions that distinguish issues addressable through documentation from those requiring deeper tooling support, improved error reporting, and ecosystem-level advances.}

\revision{Fifth, the thematic distribution is dominated by “core robotics” concerns—motion/navigation, sensing and incoming data, remote control and actuation—rather than generic programming (\textbf{RQ2}). The persistence and recent growth of these themes (\textbf{RQ3}) point to software–hardware integration boundaries as key pain points. Research should therefore prioritize frameworks, middleware, and tooling that improve interoperability, integration, and reuse across heterogeneous robotic stacks. Finally, the decline in \textit{Specification} and \textit{Task Management} questions warrants further investigation, as it may reflect reduced topic diversity, migration to other support venues, or ecosystem consolidation in hardware and software options.}


\revision{For practitioners, low passive visibility coupled with relatively high interactive engagement (\textbf{RQ1}) suggests that successful resolution often depends on targeted participation by domain-relevant contributors rather than broad exposure. A practical implication is to increase the \emph{answerability} of posts by providing reproducible context up front: hardware and firmware/middleware versions, OS/toolchain details, constraints (including timing/real-time requirements), expected versus observed behavior, minimal runnable snippets, and relevant logs, as well as what has already been attempted. Given the prevalence of clarification in comments (\textbf{RQ1}), this practice can reduce iteration cycles and time-to-resolution. In parallel, because \textit{How} questions dominate and \textit{What} questions remain substantial—particularly within \textit{Specifications} (\textbf{RQ4})—educators should prioritize instructional materials that directly target these information needs, such as step-by-step tutorials, structured FAQs, and practice-oriented guides that promote learner self-sufficiency in diagnosing and resolving real-world robotics issues.}

In future work, we aim to utilize the dedicated robotics channel within the \textsc{StackExchange} site\footnote{https://robotics.stackexchange.com (last accessed: 16-Aug-2024)}, which currently features approximately 46,500 questions. To filter out the false positives, our initial step will involve developing a large language model (LLM) based classifier to determine whether a question pertains to robotics. The LLM-based model will be provided with both positive (real robotics questions) and negative examples (non-robotics questions) to improve its precision and recall. Based on the LLM-generated filtered set of robotics questions, we will apply the Latent Dirichlet Allocation (LDA) algorithm~\cite{blei2003latent} to automatically group similar questions, allowing us to manually label each group with a meaningful theme. LDA has been widely employed in software engineering research (e.g.,~\cite{Bangash-ml:2019,Uddin:2021,Abdellatif:2020}), particularly when dealing with large volumes of posts, where manual labeling becomes impractical. The findings from this study are expected to provide deeper insights by enabling us to draw conclusions from a significantly larger dataset of robotics questions. 

Furthermore, analyzing developers' reported issues from robotics projects in open-source repositories, such as GitHub, can provide valuable insights to enhance our analysis of RQ2 and RQ4.

\section{Threats to Validity}
\label{subsec:threats}

Several threats may have impacted the findings of our paper. 

\textbf{Internal validity.}
A significant threat to our research is researcher bias and errors in identifying common themes, and categorizing the questions into \textit{What}, \textit{Why}, and \textit{How} types. This threat was mitigated because multiple authors independently labeled the questions and verified each other's labels. 
For example, a domain expert (PhD student in robotics) manually labeled 300 randomly selected questions and identified common themes, as detailed in section \ref{subsec:rq2}. Two other authors were then trained on these questions and themes before they started labeling the remaining questions. To further minimize bias, conflicts between these two authors were resolved after consulting with another author. 

\textbf{External Validity.}
\textsc{StackOverflow} is one of the most popular question-and-answer websites for computer programmers, and we used this platform to collect our data. However, despite its widespread use among developers, our findings may not be universal as we did not consider other platforms. Moreover, To minimize noise in our dataset, we specifically analyzed posts tagged with \emph{robot} and \emph{robotics}. This approach, while reducing irrelevant data, might have caused us to overlook relevant posts that did not carry these specific tags. Consequently, some questions related to robotics may have been excluded from our dataset.

%% file: 7_conclusion.tex
\section{Conclusion}
\label{sec:conclusion}

In this paper, we analyzed 500 \textsc{StackOverflow} posts related to the field of robotics. Our findings reveal a growing number of questions about robotics, although they are less popular compared to the average \textsc{StackOverflow} inquiry. We identified 11 distinct themes that robotic practitioners commonly inquire about, each with varying frequencies. Considering different popularity metrics, we found that \textit{Specifications} questions were very popular in the beginning, but the number of \textit{Specification} questions has shown a notable decline over time. Consistent themes in robotic discussions included \textit{Moving}, \textit{Incoming}, \textit{Actuator}, and \textit{Remote}. Additionally, robotic developers predominantly ask \textit{How} questions, with very few inquiries focused on the underlying reasoning (\textit{Why} questions). 

We hope that our study benefits robotic developers by highlighting the common themes they should prioritize in their learning. It should also assist educators in designing effective curricula and inform researchers about relevant topics for further exploration in the field of robotics and software engineering. 

\section{Acknowledgements}
The authors would like to thank the University Research Grant
Program (URGP) at the University of Manitoba for supporting
this research project. We also acknowledge the partial support
from the University of Manitoba Undergraduate Research Awards
(URA).

\revision{
\textbf{Declaration of generative AI and AI-assisted technologies in the manuscript preparation process.} During the preparation of this work, the authors used ChatGPT to improve readability and language (e.g., rephrasing, grammar, and clarity). The authors reviewed and edited the resulting text and take full responsibility for the content of the published article.}

\section{Data Availability} To enable verification and replication, we publicly shared our dataset\footnote{https://zenodo.org/records/13892239} that contains all the initial codes and themes produced through the thematic analysis. Additionally, we also shared the categorization of \emph{How}, \emph{Why}, and \emph{What} questions.   

\section{Conflict of Interest} The authors do not have any conflict of interest to disclose. 


